\documentclass[aps,prl,twocolumn]{revtex4-1}

\usepackage{graphicx}

\def\qr{{\bf r}}                                   
\def\qk{{\bf k}}                                   

\begin{document}

\title{Excitations and Stripe Phase Formation in a 2D Dipolar Bose Gas with Tilted Polarization}

\author{A. Macia$^1$, D. Hufnagl$^2$, F. Mazzanti$^1$, J. Boronat$^1$, R. E. Zillich$^2$}

\affiliation{$^1$ Departament de F\'{\i}sica i Enginyeria Nuclear, Campus Nord
  B4-B5, Universitat Polit\`ecnica de Catalunya, E-08034 Barcelona, Spain}
\affiliation{$^2$ Institut f\"ur Theoretische Physik, Johannes Kepler
  Universit\"at, Altenbergerstr. 69, 4040 Linz, Austria}
  

\begin{abstract}

We present calculations of the ground state and excitations of an anisotropic
dipolar Bose gas in two dimensions, realized by a non-perpendicular
polarization with respect to the system plane.  For sufficiently high density
an increase of the polarization angle leads to a density instability of 
the gas phase in the direction where the anisotropic interaction is strongest.
Using a dynamic many-body theory, we calculate the dynamic structure function
in the gas phase which shows the anisotropic dispersion of the excitations.
We find that the energy of roton excitations in the strongly interacting direction
decreases with increasing polarization angle and almost vanishes close to the instability.
Exact path integral ground state Monte Carlo simulations show that this instability
is indeed a quantum phase transition to a stripe phase, characterized by long-range
order in the strongly interacting direction.

\end{abstract}



\maketitle

Strongly correlated dipolar Bose gases in two dimensions (2D) polarized
along the direction normal to the system plane have been extensively
investigated in recent
years~\cite{goldenPRE10,astraPRL07,mazzantiPRL09,filinovPRL10}.
The ratio between the dipolar length
$r_0=mC_{dd}/(4\pi\hbar^2)$ and the average interparticle distance provides
a measure of the strength of the interaction.
$C_{dd}$ is the coupling constant proportional to the square of the (magnetic $\mu$ or
electric $d$) dipole moment, resulting in a dipolar length that can range
from a few \AA\, for magnetic dipolar systems like $^{52}$Cr
($\mu=6\mu_B$, with $\mu_B$ the  Bohr magneton), to
thousands of \AA\ for heteronuclear polar molecules
like KRb, LiCs~\cite{niPCCP09}, or RbCs~\cite{takekoshiPRA12}.
However, chemical reactions and three-body losses impose
limitations on what can be measured in
experiments with polar molecules. Therefore, recent efforts focus also on
exotic lanthanide magnetic systems like $^{164}$Dy or $^{168}$Er,~\cite{aikawaPRL12}
where the combined effect of a large magnetic moment
($\mu=10\mu_B$ for $^{164}$Dy and $\mu=7\mu_B$ for $^{168}$Er) and a
large mass, lead to dipolar length scales that, although still
significantly lower than the corresponding value for polar molecules,
is several times larger than that of $^{52}$Cr.
Er$_2$ with $\mu=14\mu_B$ and twice the mass of Er would reach even
higher values of $r_0$~\cite{ferlainoprivatecomm}.

A 2D dipolar Bose gas polarized along the normal direction to the
confining plane develops a roton excitation at high density due to the strong
repulsion between dipoles at short distances~\cite{mazzantiPRL09}.
Other works have revealed competing effects in a quasi-2D
geometry due to the head-to-tail attraction of the dipole-dipole interaction
when the third spatial dimension is added, to the point that the
system becomes unstable against density fluctuation below a critical
trapping frequency in that direction~\cite{santosPRL03,ryanPRL08,hufnaglPRL11}.
This leads to the question of 
whether a similar situation can hold in a purely 2D geometry when a
head-to-tail component to the dipole-dipole interaction is added
by tilting the polarization
with respect to the direction normal to the system plane.
The interaction becomes anisotropic,
$V(\qr) = V(x,y) = {C_{dd}\over 4\pi r^3}\Big[1-3{x^2\over r^2}\sin^2\alpha\Big]$,
with particles moving in the $x,y$-plane and a polarization
field in the $x,z$-plane, tilted by an angle $\alpha$ with 
respect to the $z$-axis. The
interaction is weakened in the $x$-direction as
$\alpha$ is increased, while it does not change in the
$y$-direction. Notice that, in the case of bosonic particles, 
only polarization angles where $V(\qr)$ is non-negative, i.e.
$\alpha \le \alpha_c = \arcsin(1/\sqrt{3})= 0.61548\dots$, are meaningful,
if there is no additional short-range
repulsion to prevent two dipoles from collapsing to a
point.

The effect of a tilted polarization on the superfluid response of
a quasi-2D dipolar Bose gas has been investigated by mean field theory~\cite{ticknorPRL11}.
The appearance of a stripe phase has been predicted in 2D dipolar Fermi systems by
approximate methods~\cite{yamaguchiPRA10,parishPRL12,sunPRB10},
observed as a spontaneous symmetry breaking even in the
isotropic case ($\alpha=0$) for high interaction strength. However,
recent fixed node diffusion Monte Carlo simulations find no evidence of that
in the isotropic case before the system crystallizes~\cite{matveevaARXIV2012}. 
In previous work~\cite{maciaPRA11} we investigated the low density
regime of the 2D dipolar Bose gas of particles interacting by the 
potential $V(\qr)$, analyzing the universal 
energy scaling properties of the anisotropic gas and  
other ground state properties.  Up to the
maximally allowed polarization angle $\alpha_c$, the low-density system 
always remains in a gaseous form and no trace of a stripe phase is found.
In this work we focus on the high density
regimes of this system, studying the effect of the anisotropy 
on the dispersion relation, especially the roton, and show how a
stripe phase forms at large densities and polarization angles.
Throughout the paper, lengths and energies are given in units of
$r_0$ and $E_0=\hbar^2/(mr_0^2)$, respectively.

\begin{figure}[t!]
\begin{center}
\includegraphics*[width=0.46\textwidth]{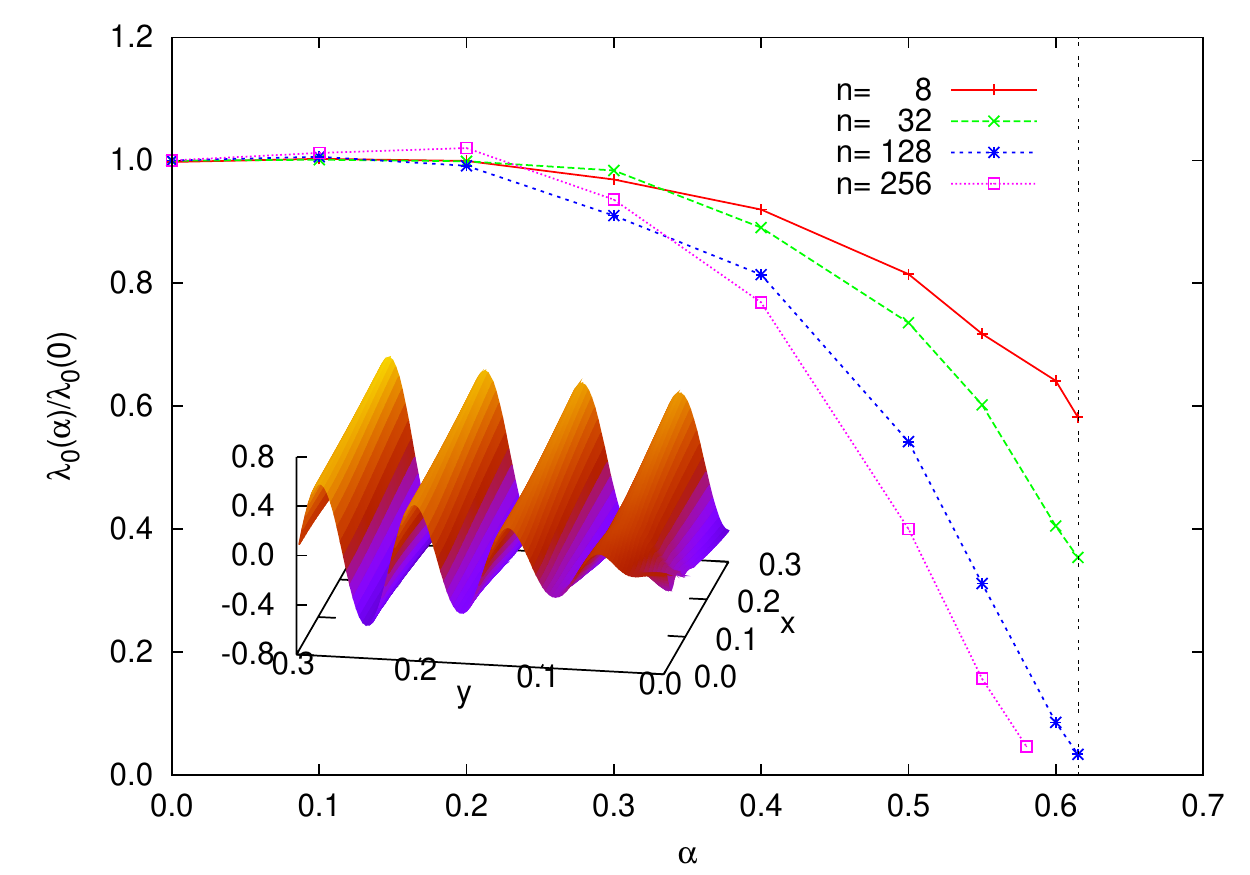}
\end{center}
\caption{
The lowest eigenvalue $\lambda_0(\alpha)$ of the hessian of the ground
state energy $E_g$ is shown in HNC-EL/0 approximation for
densities $n=8; 32; 128; 256$.  $\lambda_0(\alpha)$ is normalized by the
respective eigenvalue at $\alpha=0$.
The inset shows the eigenfunction $f_0(\qr)$ for $n=256$ and
$\alpha=0.58$.
}
\label{FIGstab}
\end{figure}

Before showing exact ground state results obtained by Monte
Carlo simulations, we present a qualitative stability analysis
of the ground state for a wide range of densities $n$ and polarization angles $\alpha$. 
We use the hyper-netted chain Euler-Lagrange
(HNC-EL)~\cite{QMBT00Polls} technique, which is based on a
Jastrow-Feenberg ansatz for the bosonic many-body wave function
$\Psi(\qr_1,\dots,\qr_N)=\exp\sum_{i<j}u_2(\qr_i-\qr_j)$.
We determine $u_2(\qr)$
variationally by imposing the extremal condition
${\delta E_g\over\delta\sqrt{g(\qr)}}=0$ for the ground state
energy $E_g$ and solving for the
pair distribution function $g(\qr)$ within the approximate HNC-EL/0 framework.
It was shown in
Ref.~\cite{castillejoPRB79} that when the lowest eigenvalue
$\lambda_0$ of the hessian $K(\qr,\qr')={\delta^2
E_g\over\delta\sqrt{g(\qr)}\delta\sqrt{g(\qr')}}$ is non-positive,
the system becomes unstable against infinitesimal
fluctuations of $g(\qr)$, where the associated eigenfunction $f_0(\qr)$
is the fluctuation driving
the instability. We have obtained the lowest eigenvalue
and eigenvector of $K(\qr,\qr')$ by imaginary-time propagation.
Figure~\ref{FIGstab} shows $\lambda_0(\alpha)$ as function of
$\alpha$ for a wide range of densities, where each curve is
normalized by the respective isotropic limit, $\lambda_0(0)$.
For all $n$, $\lambda_0(\alpha)$ decreases
with increasing $\alpha$, but there
is an important distinction between its behavior at low and high densities: for $n\lesssim
128$, $\lambda_0(\alpha)$ remains finite up to $\alpha_c$, while for
$n\gtrsim 128$ $\lambda_0(\alpha)$ falls to zero already before
reaching $\alpha_c$.
Hence the high-density gas state is unstable above a critical angle
$\alpha_0$ which is smaller than $\alpha_c$; this is also seen
by the fact that the HNC-EL equations do not converge in the
range $\alpha_0<\alpha<\alpha_c$.
The inset of Fig.~\ref{FIGstab} shows $f_0(\qr)$ for $n=256$ at the largest
angle where we found solutions to the HNC-EL equations, $\alpha=0.58$,
where $\lambda_0(\alpha_0)/\lambda_0(0)$ almost vanishes.
Away from the correlation hole at $\qr=0$,
$f_0(\qr)$ is essentially a plane wave in the more repulsive $y$-direction.
indicating a tendency of $g(\qr)$ towards long-range order in the $y$-direction.
We stress that the HNC-EL/0 results become less accurate with larger $n$,
hence the stability limits are only approximate.
Exact simulation results are presented below.

The anisotropic nature of the interaction which destabilizes the
system beyond $\alpha_0$ also influences the spectrum of elementary
excitations
which we investigate by calculating the dynamic structure function $S(\qk,E)$.
$S(\qk,E)$ is proportional to the
probability that a perturbation transfers momentum $\qk$ and energy
$E$ to the system.  Thus, for a given $\qk$, $S(\qk,E)$ has a marked
peak if $E$ coincides with the energy of an excitation of
the system. We obtain $S(\qk,E)$ using the dynamic many-body
theory~\cite{campbellPRB09}, where the equations of motion for
time-dependent fluctuations of up to pair-correlations in the
many-body wave function are solved numerically. If the convolution
approximation for the three-body distribution function~\cite{Feenberg}
is used, $S(\qk,E)$ is obtained as $S(\qk,E)=-{1\over\pi}\Im
m\Big[{S(\qk)\over E-\Sigma(\qk,E)}\Big]$, where $\Sigma(\qk,E)$ is
the complex, energy-dependent self-energy of Eq.~(2.46) in
Ref.~\cite{campbellPRB09}.
We note that the only input required to calculate $\Sigma(\qk,E)$ is the
static structure factor $S(\qk)$ of the ground state.

In order to get exact results for $S(\qk)$, we have
carried out stochastic simulations using the path
integral ground state (PIGS) Monte Carlo technique of
Ref.~\cite{sarsaJChemP00} which starts from a
variational wave function $\phi_0$ and projects out
components orthogonal to the true ground state by propagation in imaginary time.
In this sense, the result of the simulation becomes stochastically
{\em exact} provided the approximation employed for the Green's
function is accurate and the propagation time is long enough~\cite{rotaPRE10}.  In the
present case we have used as $\phi_0$ a Jastrow-Feenberg ansatz
$\prod_{i<j} f(\qr_{ij})$ built from the two-body correlation factor
$f(\qr)=K_0(2/\sqrt{r})$, corresponding to the zero-energy solution of
the two-body problem of the isotropic $1/r^3$ interaction. Despite the
isotropy of $\phi_0$, anisotropic contributions are
taken into account by a fourth-order propagator~\cite{cuervoJChemP05}, which contains the
anisotropic potential $V(\qr)$ and its
gradient. The results presented in this work have been 
obtained for $N=512$ particles in a simulation box with
periodic boundary conditions. Additionally, simulations with smaller $N$
have been carried out in order to see the $N$-dependence 
of the highest peaks in $S(\qk)$ when $n$ and $\alpha$ increase.

\begin{figure}[t!]
\begin{center}
\includegraphics*[width=0.45\textwidth]{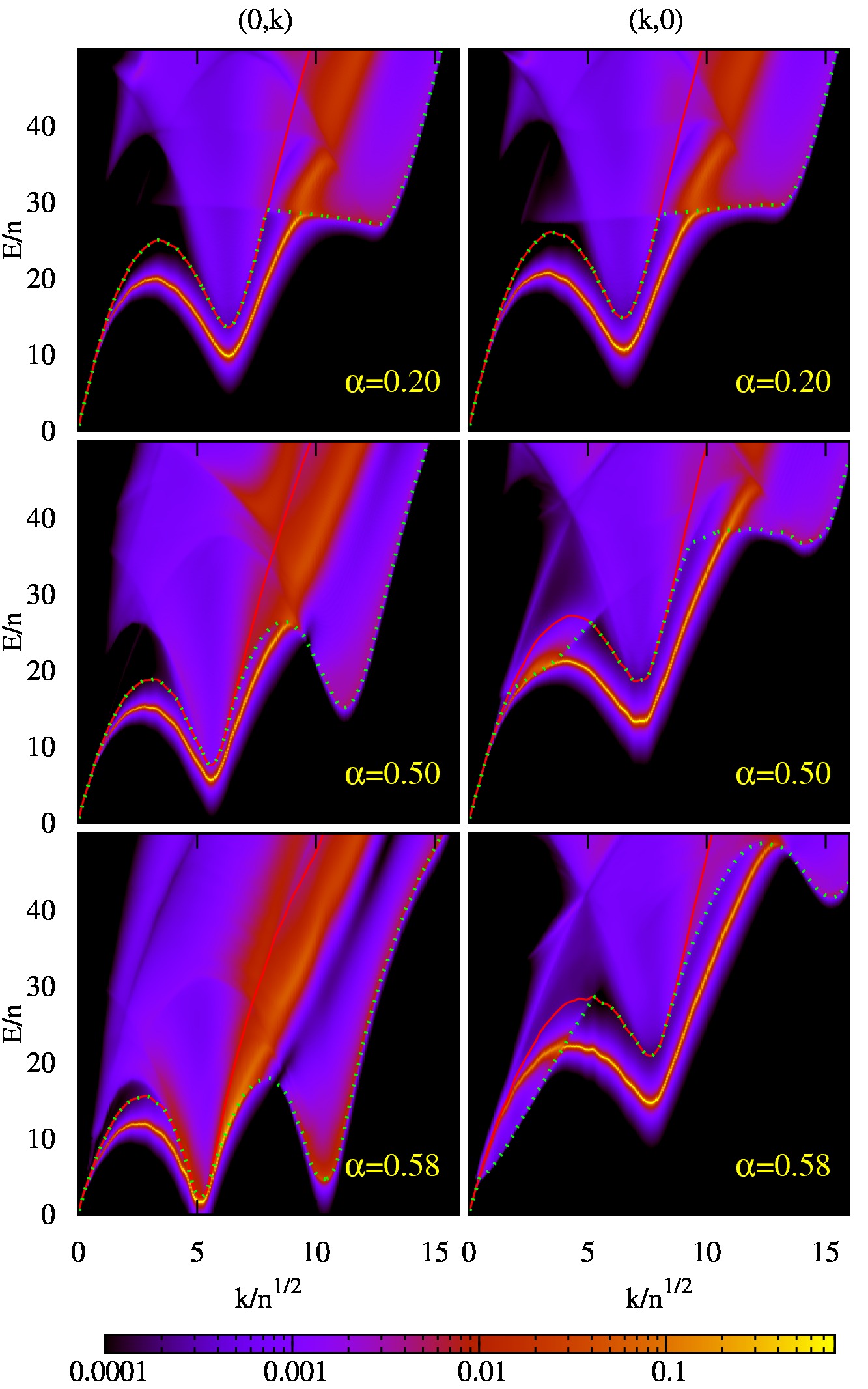}
\end{center}
\caption{
  $S(\qk,E)$ for $\qk=(0,k)$ (left panels) and $(k,0)$
  (right panels) for polarization angles $\alpha=0.20; 0.50; 0.58$ at density $n=128$.
  The spectrum in Bijl-Feynman approximation is shown as a solid line, and
  the dotted line denotes the damping limit $E_c(\qk)$.
}
\label{FIGskw}
\end{figure}

In Ref.~\onlinecite{mazzantiPRL09} we studied the density dependence
of $S(\qk,E)$ of the 2D dipolar quantum gas in the isotropic limit ($\alpha=0$).  
There we found a spectrum with a pronounced roton
for large density, due to the strong correlations induced 
by the $1/r^3$ repulsion.  Here, we are interested
in the dependence of $S(\qk,E)$ on the polarization angle $\alpha$ and,
for a given $\alpha>0$, its dependence on the direction of $\qk$.
In Fig.~\ref{FIGskw}, we show $S(\qk,E)$ for $n=128$ and
$\alpha=0.20; 0.50; 0.58$ in order to illustrate the evolution from
an isotropic to an anisotropic excitation spectrum and the approach to
the stability limit.  The wave vector
$\qk$ is pointing in the $y$ and $x$-direction (i.e. the direction of strongest
and weakest interaction) in the left and right panels.
We broaden $S(\qk,E)$ by adding a small imaginary part $\eta=0.2$
to the energy in the calculation of $\Sigma(\qk,E)$,
since otherwise undamped modes would not be visible in Fig.~\ref{FIGskw}.
Also shown is the Bijl-Feynman approximation of the spectrum,
obtained by setting $\Sigma(\qk,E)=0$ (solid line).

For $\alpha=0.20$ the dispersion is almost independent on the direction of $\qk$, with only
a slight slope of the Pitaevskii plateau~\cite{pitaevskii58}, which for isotropic systems
denotes the sudden onset of damping at twice the roton energy due to decay into two rotons.
As $\alpha$ is increased, $S(\qk,E)$ becomes very different in the $y$- and $x$-direction
and features a highly anisotropic dispersion relation for $\alpha=0.58$.
The wave number of the roton depends on the direction of $\qk$, but
most strikingly its energy decays almost to zero in the $y$-direction for $\alpha=0.58$,
indicating that the system is close to the limit where the
homogeneous gas phase in unstable against infinitesimal density fluctuations.
Since the restriction to pair correlation fluctuations
used here typically gives an upper bound to the excitation energy~\cite{campbellJLTP10},
the exact roton energy in $y$-direction is expected to be even smaller.
Furthermore, at twice the wave number of the roton, $S(\qk,E)$ has another roton-like peak
for $\alpha=0.58$, following a quadratic dispersion, albeit
broadened and with smaller spectral weight.  In the $y$-direction, the dispersion
relation thus resembles that of a solid,
continued beyond the first Brillouin zone.  While for $n=128$ and $\alpha=0.58$
the system is still in the gas phase, our PIGS results presented below
indeed predict a stripe phase at even higher density.

The dotted lines in Fig.~\ref{FIGskw} depict the damping limit $E_c(\qk)$ above which
decay into two excitations of lower energy is kinematically allowed,
hence excitations below $E_c(\qk)$ have infinite lifetime corresponding to peaks in $S(\qk,E)$
with zero linewidth.
The kinematics of an anisotropic dispersion is different from the
isotropic case, as evidenced e.g. by the lack of a constant Pitaevskii plateau.
The decay into two rotons is very efficient in an isotropic system because of
the high density of states at the roton energy.
For the anisotropic phonon-roton dispersion, the roton energy depends on the direction
of $\qk$, thus the roton energies are spread out leading to a smoother
density of states than in the isotropic limit.  For example, decay of the maxon in the
$y$-direction is not allowed, although its energy is higher than twice the roton energy.

\begin{figure}[t!]
\begin{center}
\includegraphics*[width=0.44\textwidth]{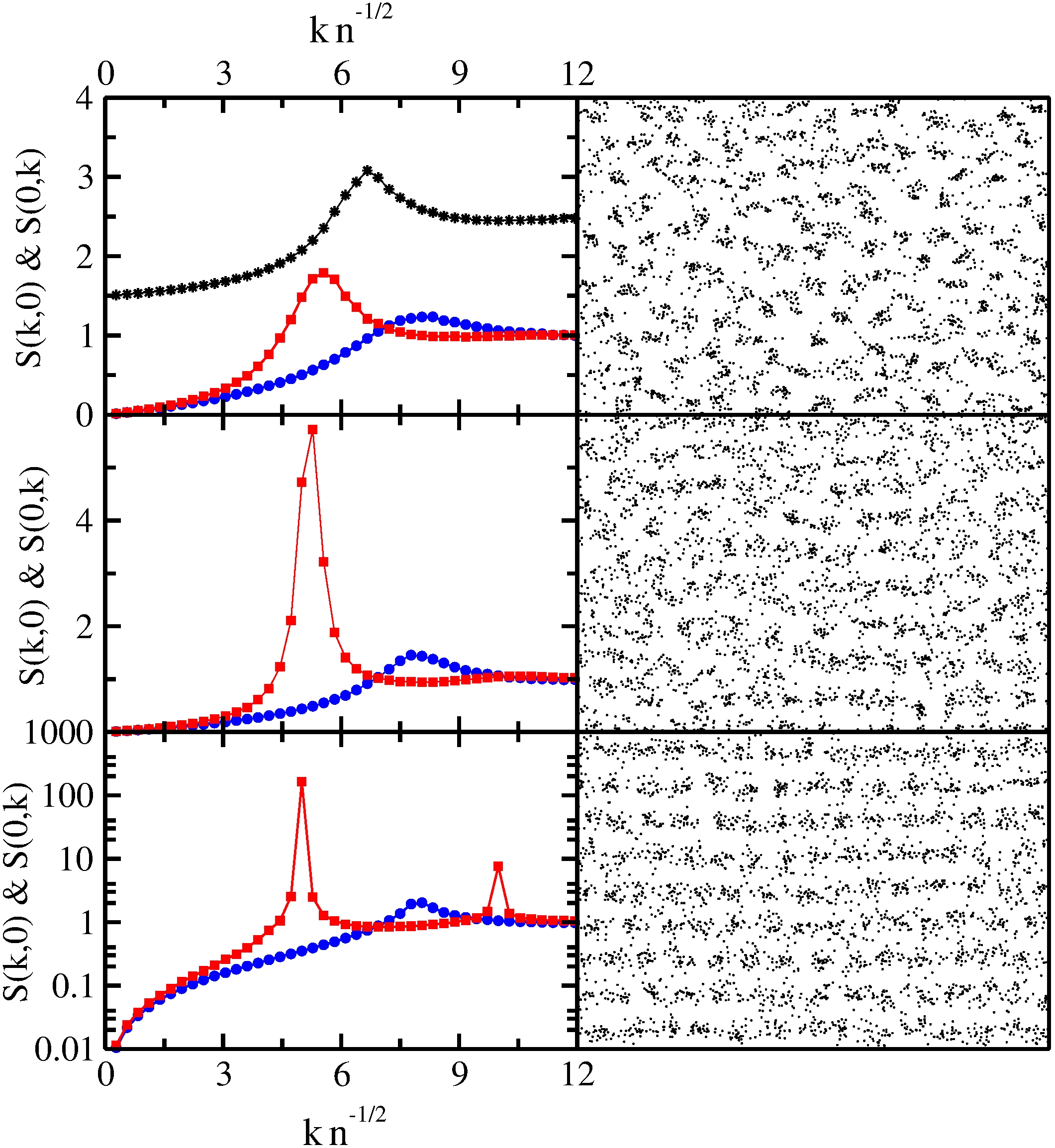}
\end{center}
\caption{
  Static structure factor (left panels) and configuration snapshots
  (right panels) at different densities and tilting angles.
  $S(k,0)$ and $S(0,k)$ for $n=64$ and $\alpha=0.58$ is shown in the top
  (blue circles and red squares, respectively). The black stars show 
  the isotropic $S(k)$ at $n=128$ and $\alpha=0$. The middle and bottom panels
  show the $S(k,0$ and $S(0,k)$ and configuration 
  snapshots for $n=128, \alpha=0.58$ and $n=256, \alpha=0.61$, respectively.
}
\label{FIGsk}
\end{figure}

Both the results for $S(\qk,E)$ and the qualitative stability analysis
(Fig.~\ref{FIGstab}) suggest that, as $\alpha$ increases,
the system develops a preference for long range order in the $y$-direction,
until the gas phase becomes unstable at a density-dependent
critical angle $\alpha_0$.  The PIGS method used to evaluate $S(\qk)$ in the
gas phase can also be used to analyze the static properties of a
system with long-range order as the present one when the homogeneous gas is not
stable anymore. As in the gas phase,
we evaluate $S(\qk)$ because long-range order can be studied by the
emergence of Bragg peaks. This is indeed what
happens when $\alpha$ is increased beyond $\alpha_0$.

Figure~\ref{FIGsk} summarizes the main PIGS results.
The upper left panel shows with black
stars the structure factor $S(k)$ (shifted up for better visibility)
for the isotropic ($\alpha=0$)
system at $n = 128$, while results along the $x$- and $y$-directions
for $n = 64$ and $\alpha=0.58$ are depicted with
blue circles and red squares, respectively.
$S(k,0)$ and $S(0,k)$ are markedly
different in the anisotropic case, which is a direct consequence of the anisotropy
of the interaction induced by the non-vanishing tilting angle.
Like in the isotropic case, the system is in the gas phase
according to our stability analysis.  We visualize
this in the upper right panel by a snapshot of
one quarter of the simulation box corresponding to the
($n = 64$, $\alpha=0.58$) case, where each worldline is
a different particle.  As expected for a gas, there is no
apparent ordering.

Results for $(n=128, \alpha=0.58)$ and $(n=256, \alpha=0.61)$
are shown in the middle and lower panel, respectively.  As can be
seen, the system becomes more anisotropic for
larger density, the peak in $S(0,k)$ is more pronounced, while
the peak in $S(k,0)$ is less affected.  For $(n=128, \alpha=0.58)$
$S(0,k)$ is still a smooth function, with a peak height that is
independent of the number of particles $N$ in the simulation.  Hence,
it is not a Bragg peak and the system is still in the gas phase.
However, for ($n=256$ and $\alpha=0.61$), the peak in $S(0,k)$ is orders of
magnitude larger than the peak in $S(k,0)$ (notice the logarithmic
scale of the bottom panel). The corresponding snapshot shows clearly the
formation of a stripe phase, which according to $S(\qk)$
is like a gas in the $x$-direction where the interaction is
weak, and a solid along the $y$-axis where the interaction is strong.
The peak in $S(0,k)$ grows almost linearly with $N$, which further supports its interpretation
as a Bragg peak.
A second peak of less but still significant
intensity develops at twice the wave number of the first peak.
We note that for $\alpha=0$ a stripe phase has not been observed and the
isotropic system remains in the gas phase until it solidifies at high density~\cite{astraPRL07}.
The same conclusion for a fully isotropic interaction has been
reported recently for a dipolar Fermi gas in 2D~\cite{matveevaARXIV2012}.

\begin{figure}[t!]
\begin{center}
\includegraphics*[width=0.49\textwidth]{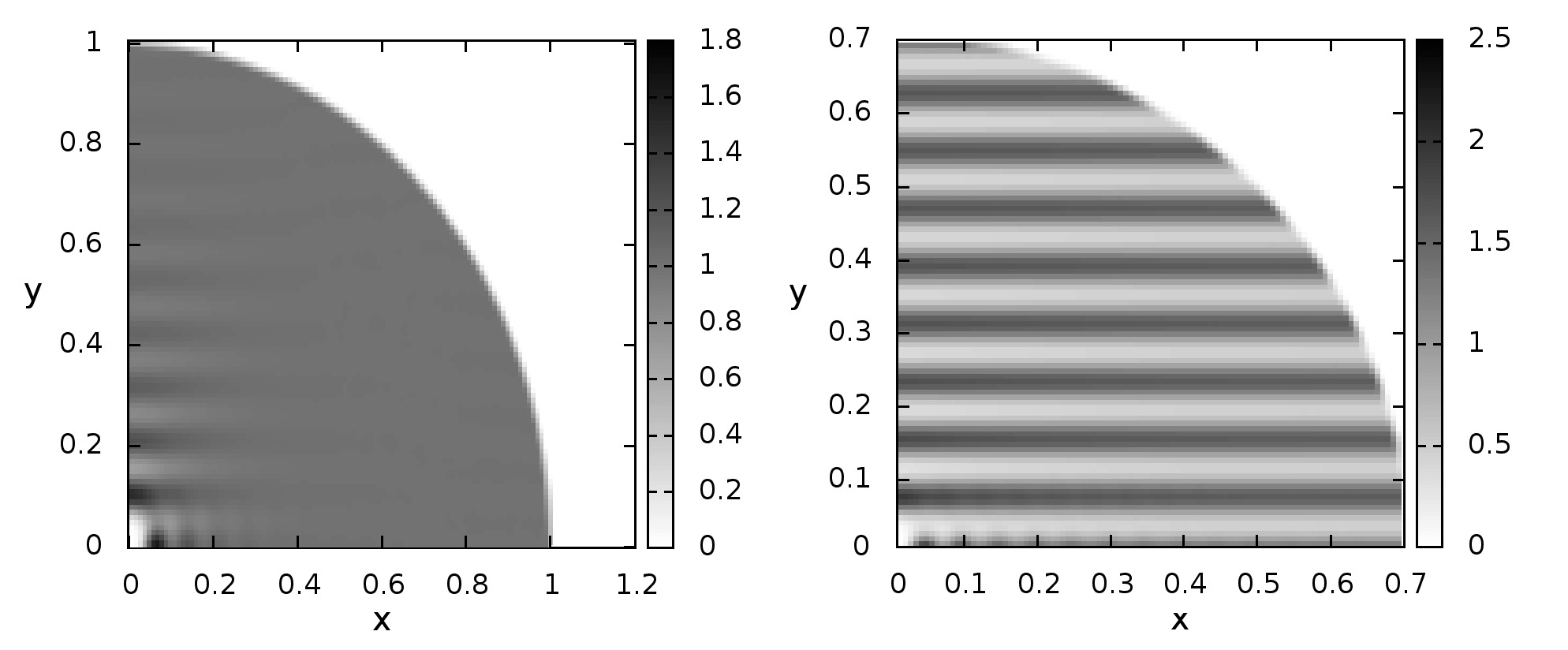}
\end{center}
\caption{
  Pair distribution function $g(x,y)$ for density $n=128$ and 
  $\alpha=0.58$ (left), and for $n=256$ and $\alpha=0.61$ (right).
}
\label{FIGg}
\end{figure}

We close the discussion showing a 3D-map of the pair
distribution function $g(\qr)$ for $|\qr|<L_-/2$ 
($L_-$ is the smaller side of the simulation box) in Fig.~\ref{FIGg} for the two cases
($n=128, \alpha=0.58$) (left panel) and ($n=256, \alpha=0.61$) (right panel). 
The stripe phase becomes clearly visible in the second case as a plane wave in the
$y$-direction, filling the whole simulation box.
For ($n=128, \alpha=0.58$) oscillations in the $y$-direction are present
only for small $x$ and are damped with increasing $y$, hence
$g(\qr)$ becomes isotropic for large $|\qr|$ and
equal to unity, consistent with the behavior of a gas.

Summarizing, we have analyzed the behavior of an anisotropic dipolar Bose gas in 2D,
using several methods: a qualitative stability analysis of the ground state
based on the HNC-EL method, the exact calculation of structural quantities from
PIGS Monte Carlo simulations, and the dynamic structure function in the
pair fluctuation approximation of the dynamic many-body theory.
All results show that for large tilting angle $\alpha$
and large density $n$, there is a quantum phase transition
to a stripe phase, characterized by long-range order in the direction of
stronger interaction.  The phonon-roton dispersion is very anisotropic in the
gas phase close to the stripe phase transition, with an almost vanishing roton energy in
the $y$-direction.

\begin{acknowledgments}
This work has been supported by the Austrian Science Foundation FWF 
under grant No.~23535, and by Grant No.~FIS2011-25275
from DGI (Spain) and Grant No.~2009-SGR1003 from the Generalitat de
Catalunya (Spain).
D.H. and R.E.Z are grateful to E. Krotscheck for helpful discussions
\end{acknowledgments}


%

\end{document}